\documentclass[%
 reprint,
 twocolumn,
 amsmath,amssymb,
aip,jcp,
]{revtex4-2}

\usepackage{natbib}
\usepackage{graphicx}
\usepackage{dcolumn}
\usepackage{bm}
\usepackage{xcolor}
\usepackage{physics}
\usepackage{lipsum}
\usepackage{placeins}

\begin{document}
\title{Unraveling Exciton Trap Dynamics and Nonradiative Loss Pathways in Quantum Dots via Atomistic Simulations}
\author{Bokang Hou}
\affiliation {Department of Chemistry, University of
California, Berkeley, California 94720, United States;}
\author{Salvatore Gatto}
\affiliation {Institute of Physics, University of Freiburg, Hermann-Herder-Straße 3, 79104 Freiburg, Germany;}
\author{Samuel L. Rudge}
\affiliation {Institute of Physics, University of Freiburg, Hermann-Herder-Straße 3, 79104 Freiburg, Germany;}
\author{Johan E. Runeson}
\affiliation {Institute of Physics, University of Freiburg, Hermann-Herder-Straße 3, 79104 Freiburg, Germany;}
\author{Michael Thoss}
\affiliation {Institute of Physics, University of Freiburg, Hermann-Herder-Straße 3, 79104 Freiburg, Germany;}
\author{Eran Rabani}
\affiliation {Department of Chemistry, University of
California, Berkeley, California 94720, United States;}
\affiliation {Materials Sciences Division, Lawrence Berkeley National Laboratory, Berkeley, California 94720, United States;}
\affiliation {The Raymond and Beverly Sackler Center of Computational Molecular and Materials Science, Tel Aviv University, Tel Aviv 69978, Israel;}

\begin{abstract} 
Surface defects in colloidal quantum dots are a major source of nonradiative losses, yet the microscopic mechanisms underlying exciton trapping and recombination remain elusive. Here, we develop a model Hamiltonian based on atomistic electronic calculations to investigate exciton dynamics in CdSe/CdS core/shell QDs containing a single hole trap introduced by an unpassivated sulfur atom. By systematically varying the defect depth and reorganization energy, we uncover how defect-induced excitonic states mediate energy relaxation pathways. Our simulations reveal that a single localized defect can induce a rich spectrum of excitonic states, leading to multiple dynamical regimes, from slow, energetically off-resonant trapping to fast, cascaded relaxation through in-gap defect states. Crucially, we quantify how defect-induced polaron shifts and exciton-phonon couplings govern the balance between efficient radiative emission and rapid nonradiative decay. These insights clarify the microscopic origin of defect-assisted loss channels and suggest pathways for tailoring QD optoelectronic properties via surface and defect engineering.
\end{abstract}

\maketitle

Colloidal semiconductor nanocrystal (NC) quantum dots (QDs) have gained significant interest in both fundamental research and technological applications owing to their size-dependent optical and electronic properties.\cite{kagan2016building,liu2021colloidal,kim2024recent} Quantum confinement in these NCs results in discrete energy levels for electrons and holes, creating quantized excitonic states with tunable absorption and photoluminescence (PL) characteristics.\cite{brus1984electron,murray1993synthesis} Although these properties have facilitated the development of QD-based devices like LEDs, lasers, and solar cells,\cite{park2021colloidal,lee2022bright,carey2015colloidal} a persistent challenge remains in enhancing photon emission efficiency or the photoluminescence quantum yield (PLQY) in such devices~\cite{hanifi2019redefining}, which is often limited by the presence of surface defect/trap states.\cite{orfield2015correlation, giansante_surface_2017, baek2024insights,steenbock2024surface}

Different types of defect states can be formed during the synthesis of colloidal NCs, where the lack of ligand coordination, surface oxidation, atomic vacancies, and stoichiometry are known to introduce localized states at the NC surface.\cite{houtepen_origin_2017,brawand_defect_2017,goldzak2021colloidal} These surface states trap either an electron or a hole, depending on the chemical environment of the surface atom and depending on the ligand passivation.\cite{giansante_surface_2017,houtepen_origin_2017,yan2020uncovering} Those traps states can be highly dynamical~\cite{utterback2016observation,cline2018nature} and strongly affect carrier dynamics in NCs and NC-based devices.\cite{allan2009fast,delerue2013nanostructures,bozyigit2013quantification} The surface states are usually viewed as detrimental to the quantum yield as a result of two central processes. First, intrinsic electrons/holes can be trapped by relaxing to the localized surface defect state(s).\cite{klimov2000optical,underwood2001ultrafast,sewall_state-resolved_2008,berthe2008probing,mcguire2010spectroscopic} The trapped electron or hole then recombines radiatively on longer timescales.\cite{hasselbarth1993detection,norris1994measurement} The broad red-shifted peak observed in the PL spectrum of poorly passivated QDs is typically assigned to the emission of the traps.\cite{chestnoy1986luminescence,mooney_challenge_2013,mooney_microscopic_2013} Second, the nonradiative recombination channel, where the recombined electron-hole pair dissipates its energy through lattice vibrations, is often much faster compared to the intrinsic excitonic states of the NC. This process is commonly termed Shockley-Read-Hall (SRH) recombination in bulk semiconductors, negatively affecting the performance of solar cells.\cite{read1950dislocation,hall1952electron,pevere_rapid_2018} 

\begin{figure*}[t]
    \centering
    \includegraphics[width=0.65\linewidth]{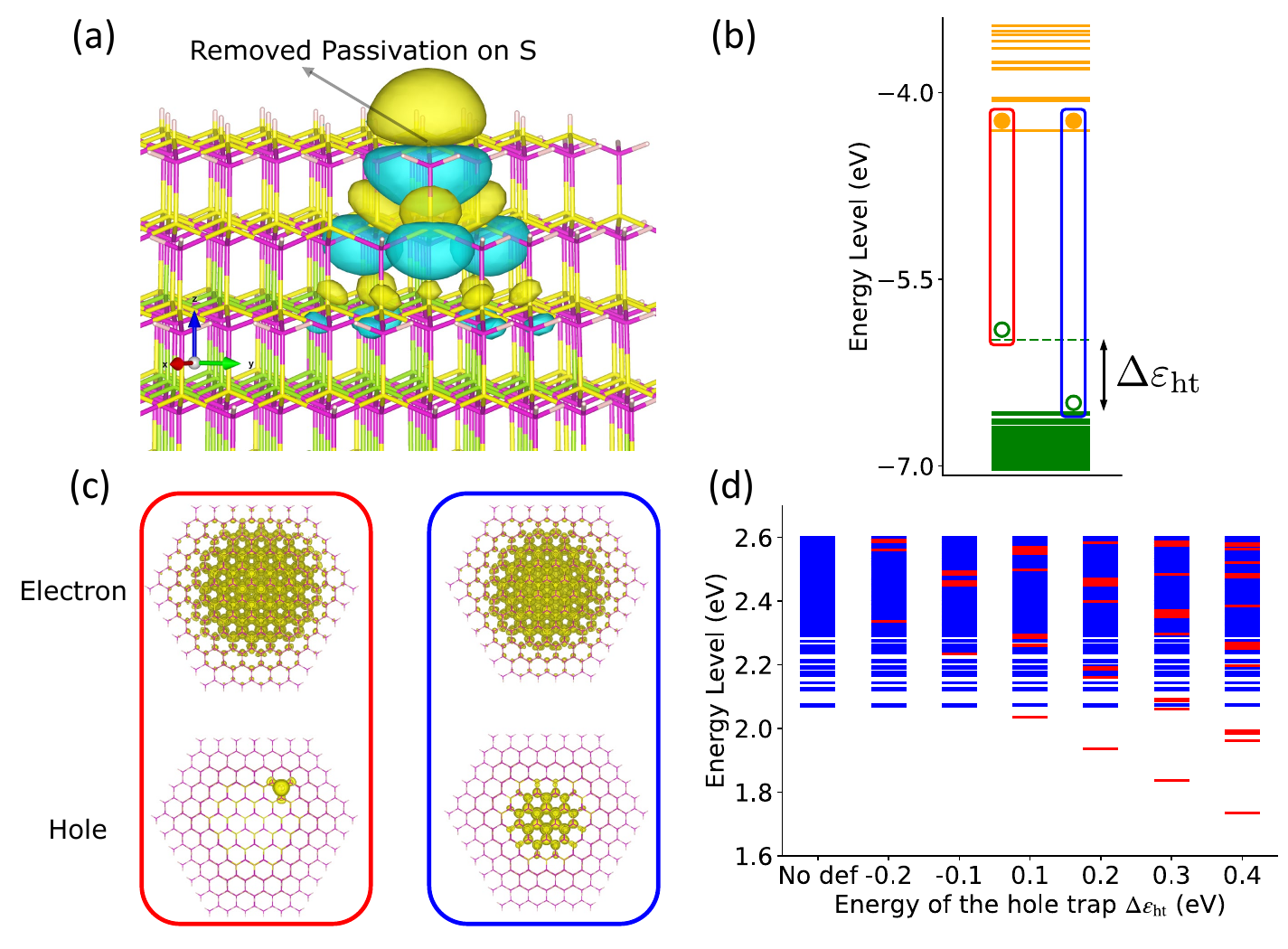}
    \caption{(a) Wavefunction isosurface of the localized hole‑trap state arising from an unpassivated surface sulfur atom. Yellow and blue indicate the positive and negative phases of the wavefunction, respectively. Cadmium (Cd), sulfur (S), and selenium (Se) atoms are rendered in magenta, yellow, and green. (b) Single‑particle energy levels of quasi‑electron (orange) and quasi‑hole (green) states from semiempirical pseudopotential calculations. The hole‑trap level is marked by a dashed line, and the trap depth is defined as $\Delta\varepsilon_{\rm ht}$. (c) Interacting electron (top) and hole (bottom) densities for a localized defect state (red box) and the lowest bright excitonic state (blue box) from the Bethe–Salpeter equation (BSE) calculation. Note that the electron density shifted slightly towards the hole trap density in the defect state. (d) Energy levels of the interacting intrinsic excitonic (blue) and defect states (red) from shallow ($-$0.2~eV) to deep (0.4~eV) single-particle hole trap depth $\Delta\varepsilon_{\rm ht}$. Defect levels shift downward with increasing $\Delta\varepsilon_{\rm ht}$.}
    \label{fig:intro_fig}
\end{figure*}

In the interacting electron-hole pair or excitonic framework, both intrinsic excitonic states (states without trapped electron or hole states) and defect excitonic states can be treated on the same footing. The intrinsic excitons relax via defect excitons by emitting one or more phonons before ultimately recombining back to the ground state (either radiatively or nonradiatively).  Both experimental and theoretical studies have shown that the passivation ligands significantly alter the electronic properties and reorganization energy of defect states.\cite{jones_quantitative_2009,mooney_microscopic_2013,liu2015ligands,houtepen_origin_2017,alexander2024understanding} However, it remains challenging to describe how changes in these defect properties translate to the relaxation dynamics of excitons. This is because such a correlation necessitates modeling the process at an atomistic level, where the intricate interactions between the ligands, the surface, and the defect states must be accurately captured to understand their impact on excitonic behavior.

In this paper, we explore the exciton dynamics in the presence of surface defects in CdSe/CdS core/shell colloidal QDs. An atomistic model was developed to include the interacting electrons and trapped holes (or defect exciton), as well as the coupling to the lattice vibrations of the QD. We focus on two dynamical processes, exciton trapping and defect-assisted nonradiative recombination. We demonstrate how the depth of hole traps and the reorganization energy of defect excitons influence the relaxation dynamics of excitons and the resulting quantum yield, in an unexpected way. By systematically analyzing the exciton relaxation pathways, we identify distinct regimes of defect-induced phenomena. Our findings provide valuable insights into the underlying mechanisms that govern defect-related behavior in NC QDs, offering guidance for the design of defect-tolerant materials and devices. 

To investigate the exciton relaxation and trapping processes, it is essential to explicitly incorporate the energies and nuclear couplings of both the excitonic and defect states into the model.~\cite{prezhdo2009photoinduced,jasrasaria2022simulations} Our previous research has focused on developing pseudopotentials~\cite{cohen1967,wang1995local,fu_inp_1997} (PPs) for intrinsic excitons that accurately capture electron and hole energies and their coupling to the lattice. This approach allowed us to theoretically describe a wide range of carrier dynamics, including hot exciton cooling, photoluminescence spectra, and electron relaxation and transfer.\cite{jasrasaria2022simulations,jasrasaria2021interplay, jasrasaria2023circumventing, lin_theory_2023, hou2023incoherent,coley2024intrinsically} Here, we will use a similar approach to characterize the defect excitons and their impact on excitonic dynamics and quantum yield.  However, since our PPs were parameterized for bulk materials, we will treat the electron/hole trap energy and the overall reorganization energy of the trap parametrically, as further discussed below.

\begin{figure*}[t]
    \centering 
    \includegraphics[width=0.9\linewidth]{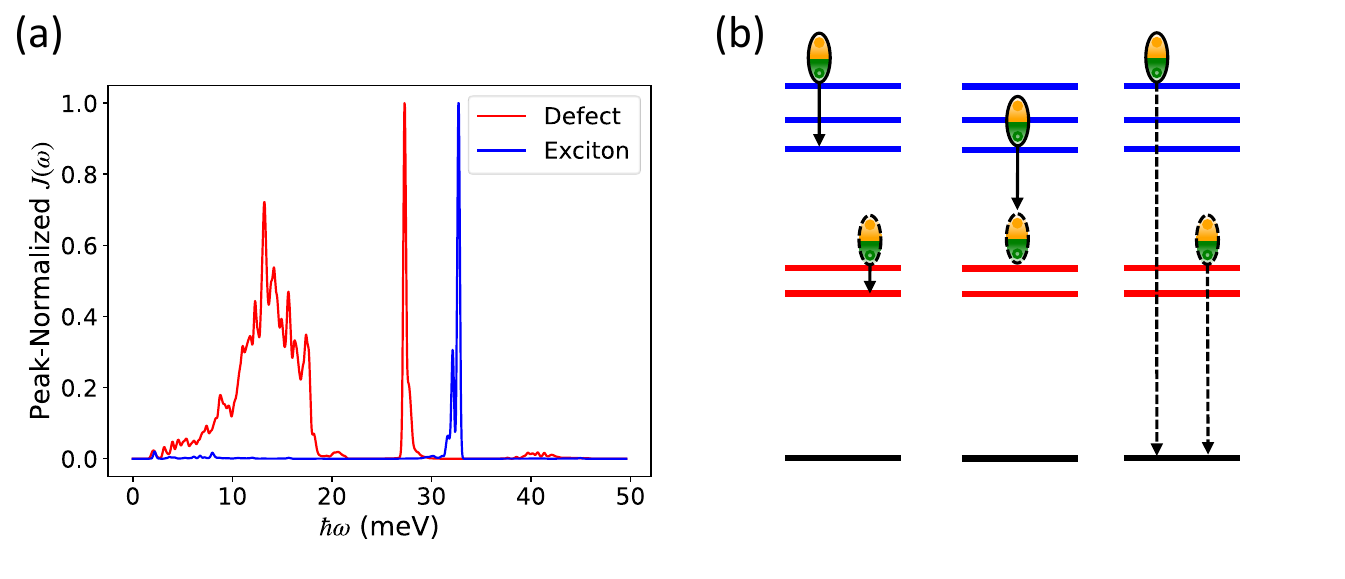}
    \caption{(a) Peak-normalized diagonal spectral density for the lowest intrinsic excitonic and defect state. (b) Schematic illustration of the exciton dynamics captured by the model Hamiltonian: from left to right, exciton–defect relaxation, exciton trapping, and nonradiative recombination of excitons or defect states. }
    \label{fig:ext_def_spectral}
\end{figure*}

We begin by describing the model Hamiltonian, $H_{\rm QD}$,  used to generate the excitonic dynamics, followed by an explanation of how it was parameterized for CdSe/CdS core-shell nanocrystals with a single defect surface. The model Hamiltonian is expressed as follows:\cite{jasrasaria2023circumventing}
 \begin{align}
 \begin{aligned}
    & H_{\rm QD} = E_g\ket{\psi_g}\bra{\psi_g} +\sum_{\alpha}\left(\frac{1}{2}{P}_{\alpha}^2 + \frac{1}{2}\omega_\alpha^2Q_\alpha^2\right) \\
    & + \sum_{n, m\in \rm{exct}} \left(E_{e_n}^0\delta_{nm} + \sum_\alpha V_{e_ne_m}^\alpha Q_\alpha\right)\ket{\psi_{e_n}}\bra{\psi_{e_m}}\\
    & + \sum_{n, m\in \text{def}}\left(E_{d_n}^{0}\delta_{nm}+\sum_{\alpha}\mathcal{V}_{d_nd_m}^\alpha{Q}_{\alpha}\right)\ket{\psi_{d_n}}\bra{\psi_{d_m}}\\
    & +  \sum_{\substack{n\in\text{def}, \\m\in\text{exct}\\\alpha}} \ket{\psi_{d_n}}\bra{\psi_{e_m}}\mathcal{V}_{d_ne_m}^\alpha{Q}_{\alpha} + \sum_{\substack{n\in\text{def}\\\alpha}} \ket{\psi_{d_n}}\bra{\psi_g}\mathcal{V}_{{d_n}g}^\alpha{Q}_{\alpha} \\
    &+ \sum_{\substack{n\in\text{exct}\\\alpha}}\ket{\psi_{e_n}}\bra{\psi_{g}} V_{e_ng}^\alpha Q_\alpha + h.c.
 \end{aligned}
 \label{eq:hamiltonian}
\end{align}
Here \(E_g\) is the ground‐state energy of \(\ket{\psi_g}\).  The index \(n=1,\dots,N_{\rm exct}\) labels the intrinsic‐exciton manifold \(\{\ket{\psi_{e_n}}\}\), while \(n=1,\dots,N_{\rm def}\) labels the defect‐exciton manifold \(\{\ket{\psi_{d_n}}\}\). The energies of the intrinsic excitons $E_{e_n}^0$ and the defect excitons $E_{d_n}^0$ were computed in the equilibrium geometry (the superscript 0 indicates $\mathbf{Q} = 0$). $h.c.$ denotes the Hermitian conjugate of all preceding off‐diagonal terms. The model Hamiltonian is written in the single-exciton basis, where the multi-exciton effects are excluded. The nuclear degrees of freedom of the QD were described using normal mode coordinates ($Q_\alpha$) and momenta ($P_\alpha$), each characterized by a normal mode frequency ($\omega_\alpha$). These were obtained from a suitable force field,\cite{zhou_stillinger-weber_2013}, accounting for intrinsic, surface, and defect modes. The remaining terms in the Hamiltonian describe the exciton-phonon coupling strengths. $V_{e_ne_m}^{\alpha}$ represents the coupling between intrinsic exciton states, while $\mathcal{V}_{d_nd_m}^{\alpha}$ denotes couplings that involve at least one defect exciton state.

Experimental and theoretical studies have shown that dangling chalcogenide atoms lead to the formation of hole-trapping excitonic states within the band gap of the CdSe–CdS core–shell quantum dots.\cite{sowers2016photophysical,pu2016battle,puzder2004self,hesitu} In the PP calculation, fully passivated structures yield clean band gaps, where the ligand potential shifts the surface electron and hole states into the respective band.  To model the trap state, we removed one passivation ligand from a sulfur surface atom along the [001] crystal plane, and then generated electron and hole states near the band edge using the same PP approach.\cite{fu1997local,jasrasaria2020sub}  In addition to the intrinsic electron and hole states, a single localized quasi-hole trap state emerged, with energy difference $\Delta\varepsilon_{\rm ht}=\varepsilon_{\rm ht}-\varepsilon_{\rm hole}$ from the highest hole state. The manifold of electron-, hole- and trap-states was used to solve the Bethe–Salpeter equation (BSE),\cite{rohlfing2000electron,eshet2013electronic} with the trap energy ($\varepsilon_{\rm ht}$) treated parametrically by varying its depths. 

The existence of this hole trap state has been the subject of numerous studies.\cite{wuister2004influence,califano_temperature_2005,califano2013universal,veamatahau2015origin,jasrasaria2020sub} As illustrated in Fig.~\ref{fig:intro_fig}(a), the hole wavefunction, obtained from the PP calculations, exhibits an sp$^3$-like symmetry~\cite{shu2015defect,peng2018dynamics} and is localized near the unpassivated S atom, within the CdS shell. The energy levels of single-particle electron and hole states near the band gap are plotted in Fig.~\ref{fig:intro_fig}(b). The green dashed line represents the energy level of the trapped hole. For $\Delta\varepsilon_{\rm ht}>0$, the hole trap lies above the highest intrinsic hole state, essentially in the middle of the fundamental gap. In this case, $\Delta\varepsilon_{\rm ht}$ indicates the depth of the hole trap within the band gap.
Conversely, when $\Delta\varepsilon_{\rm ht}$ is negative, the hole trap resides within the valence band. 
 
In the BSE, excitonic states are represented as linear combinations of electron and hole pair-states. In the absence of a hole trap, the lowest excitonic states are predominantly composed of band-edge quasi-electron and quasi-hole states, as illustrated in Fig.~\ref{fig:intro_fig}(c). The projected density of the lowest exciton for a fully passivated QD is shown in the blue box, where the hole remains localized to the core while the electron extends into the shell, reflecting quasi-type II behavior.\cite{eshet2013electronic} 

In the presence of a hole trap state with an offset of $\Delta\varepsilon_{\rm ht}>0$, the lowest excitonic state primarily consists of the trapped hole and the lowest intrinsic electron state, forming a trapped exciton or a defect excitonic state. In this case, the projected density shown in the red box of panel (c) reveals that the interacting hole is localized near the defect center (as in the quasi-hole trap), and the interacting electron density is slightly shifted towards the localized hole because of Coulomb attraction. The excitonic energies of the CdSe/CdS core-shell QD are plotted in Fig.~\ref{fig:intro_fig}(d). The energy of the defect excitons (shown in red) decreases as $\Delta\varepsilon_{\rm ht}$ increases. Notably, a single trap state may generate multiple excitonic defect states below the energy of the lowest intrinsic exciton, depending on the depth of the hole trap.

\begin{figure*}[t]
    \centering
    \includegraphics[width=0.8\linewidth]{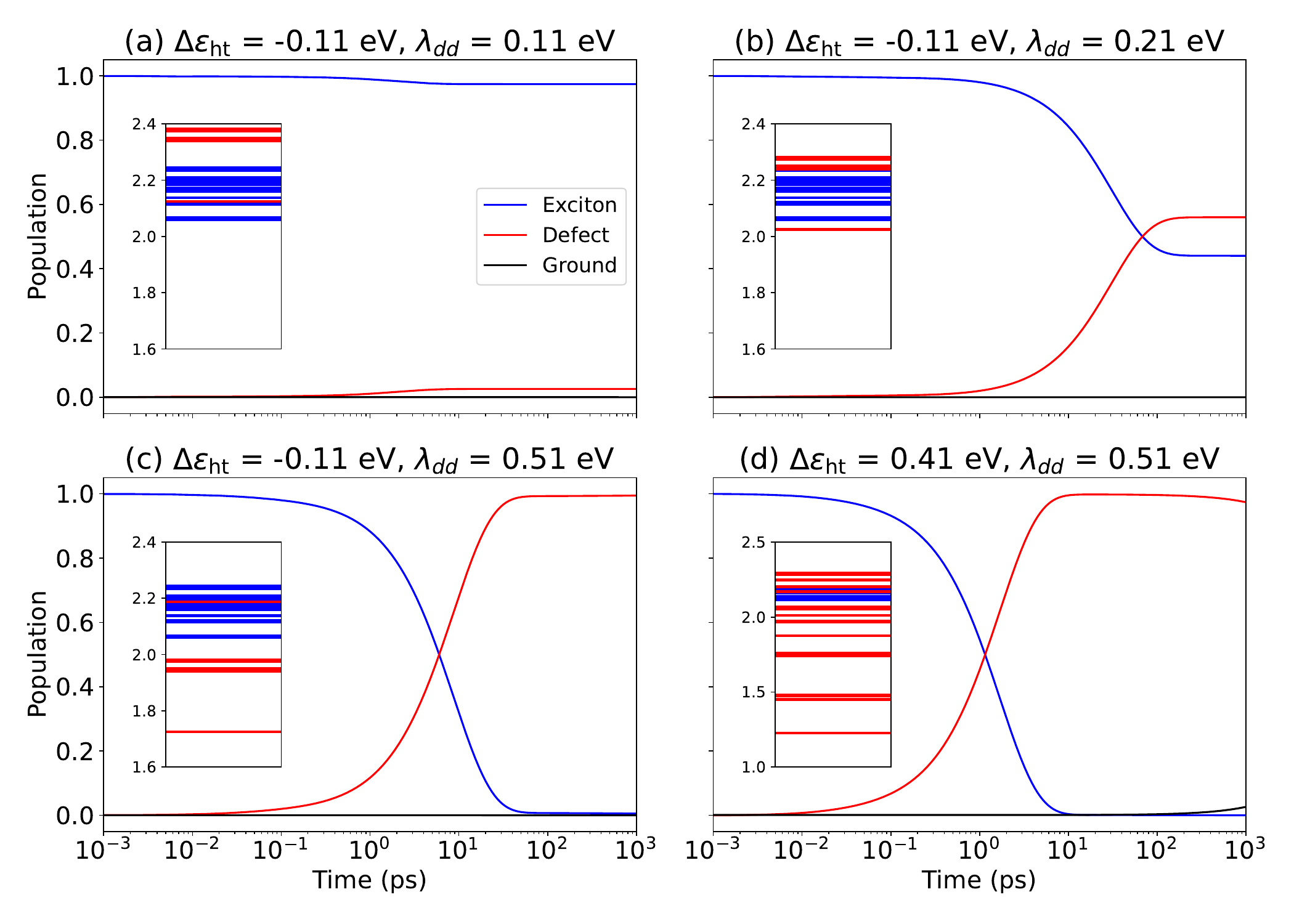}
    \caption{Multi‐defect population dynamics for varying hole‐trap depths $\Delta\varepsilon_{\rm ht}$ and defect reorganization energies $\lambda_{dd}$. Panels (a)--(d) correspond to:  
    (a) $\Delta\varepsilon_{\rm ht}=-0.11\,\mathrm{eV}$, $\lambda_{dd}=0.11\,\mathrm{eV}$;  
    (b) $\Delta\varepsilon_{\rm ht}=-0.11\,\mathrm{eV}$, $\lambda_{dd}=0.21\,\mathrm{eV}$;  
    (c) $\Delta\varepsilon_{\rm ht}=-0.11\,\mathrm{eV}$, $\lambda_{dd}=0.51\,\mathrm{eV}$;  
    (d) $\Delta\varepsilon_{\rm ht}=0.41\,\mathrm{eV}$, $\lambda_{dd}=0.51\,\mathrm{eV}$.  
    An off‐diagonal reorganization energy of $\lambda_{ed}=0.1\,\mathrm{meV}$ quantifies the coupling between the defect and the intrinsic excitonic states. The total populations of excitons, defects, and ground states are plotted in blue, red, and black, respectively; intrinsic and defect populations are summed over all individual states. Insets show the polaron energy levels $\mathcal{E}_n = E_n^0 - \lambda_{nn}$ (in eV) for the intrinsic excitons (blue) and defect states (red) involved in the dynamics.}
    \label{fig:multi_defect_population}
\end{figure*}

The final component necessary to describe the excitonic dynamics is the exciton-phonon coupling, characterized by the spectral density, $J_{nm}(\omega)$:
\begin{align}
\begin{aligned}
    J_{nm}(\omega) &= \pi\sum_\alpha\omega_\alpha\lambda^\alpha_{nm}\delta(\omega-\omega_\alpha)
\end{aligned}
\end{align}
where the mode-wise reorganization energy is defined as:
\begin{equation}
\begin{split}
\lambda_{e_ne_m}^\alpha = \frac{1}{2}(V_{e_ne_m}^\alpha/\omega_\alpha)^2 \\
\lambda_{d_nd_m}^\alpha = \frac{1}{2}(\mathcal{V}_{d_nd_m}^\alpha/\omega_\alpha)^2
\end{split}
\end{equation}
for intrinsic excitons and defect excitons, respectively. The diagonal spectral density $J_{nn}(\omega)$ corresponds to the energy change between the equilibrium geometry of an excitonic state relative to the ground state, while the off-diagonal spectral density, $J_{nm}(\omega)$, represents the magnitude of the nonadiabatic coupling between different excitonic states via lattice vibration with phonon frequency $\omega$.

Fig.~\ref{fig:ext_def_spectral}(a) shows the diagonal spectral densities for the lowest intrinsic and defect exciton state (normalized to the peak), for $\Delta\varepsilon_{\rm ht}>0$, calculated from the PP model. As observed previously,\cite{salvador2006exciton,lin2015electron,lin_theory_2023} intrinsic excitons couple strongly to the localized CdSe core vibrations around 33~meV. The enhanced coupling to the optical mode explains the prominent phonon sideband observed in both the calculated and measured PL spectra.\cite{lin_theory_2023,yazdani2020size} Defect excitons, on the other hand, couple strongly to a broad band of lattice modes in the 8-16 meV range, including the breathing and torsional modes of the QD. The sharp peak at 29 meV corresponds to the coupling of the defect state to the local surface modes on the [001] plane where the defect resides.

The magnitude of the defect coupling $\mathcal{V}_{nm}^\alpha$ may depend specifically on the type of ligand and surface reconstruction.\cite{bozyigit2016soft,srivastava2017understanding,yazdani2020size,guzelturk2021dynamic} This is hard to model even with \textit{ab initio} methods.\cite{brawand_defect_2017,jin2021photoluminescence,dai2024excitonic} We therefore, treat the reorganization energy of the defect as a parameter (see SI for the procedure to scale $\mathcal{V}_{d_nd_m}^\alpha$). In particular, all diagonal couplings for the defect states, $\mathcal{V}_{d_nd_n}^{\alpha}$, are scaled relative to the lowest-energy defect state, whose reorganization energy is defined as $\lambda_{dd}=\frac{1}{2}\sum_\alpha (\mathcal{V}_{d_1d_1}^\alpha/\omega_\alpha)^2$. Since the defects experience larger polaron shifts compared to intrinsic excitons due to their localized nature,\cite{mooney_challenge_2013,mooney_microscopic_2013} we consider a range of $\lambda_{dd}$ from $100$ to $600$~meV. These values are more than an order of magnitude larger than the values calculated for intrinsic excitons ($10$-$20$~meV for CdSe/CdS core/shell QDs). In contrast, the off-diagonal reorganization energies between localized defect and intrinsic excitons, defined as $\lambda_{ed}=\frac{1}{2}\sum_\alpha (\mathcal{V}_{e_1d_1}^\alpha/\omega_\alpha)^2$, are found to be smaller ($0.1$–$0.5$~meV), reflecting the negligible spatial overlap between the corresponding excitonic wavefunctions. In the regime where $\lambda_{ed} \ll \lambda_{dd}$ and $\lambda_{ed} <\lambda_{ee}$, the trapping rates scale with the square of the coupling between the defect and intrinsic excitons, $\mathcal{V}_{ed}^2$. Therefore, the specific value of $\lambda_{ed}$ affects the dynamics in a predictable way, so we uniformly set it to $0.1$~meV for all defect states.


The dynamics generated by the model Hamiltonian were computed using a quantum master equation after applying a small polaron transformation to the Hamiltonian and expanding the propagator up to second order in the dressed exciton-phonon coupling (further details are provided in the Supporting Information). The validity of this polaron-based perturbative approach relies on diagonal reorganization energies being larger than the off-diagonal ones. We observe that this condition is satisfied for intrinsic excitons ($\lambda_{e_ne_n} > \lambda_{e_ne_m}$) and even more strongly so for defect excitons ($\lambda_{d_nd_n} \gg \lambda_{d_nd_m}$).
The time evolution of the polaron-transformed reduced density matrix within the Hilbert space of intrinsic and defect excitons, $\tilde{\sigma}_{nm}(t) = \bra{\psi_n}\tilde{\sigma}(t)\ket{\psi_m}$, is given by~\cite{nitzan2024chemical}
\begin{align}\label{eq:pt_eom}
\frac{\partial \tilde{\sigma}_{nm}(t)}{\partial t} 
&= -i \tilde{\omega}_{nm}\,\tilde{\sigma}_{nm}(t) + \sum_{kl} K_{nm,kl}(t)\,\tilde{\sigma}_{kl}(t).
\end{align}
Here, the normalized energy gap frequency is given by $\tilde{\omega}_{nm} = (\mathcal{E}_n - \mathcal{E}_m)/\hbar = (E_n^0 - \lambda_{nn} - E_m^0 + \lambda_{mm})/\hbar$. We observe that the population dynamics $p_n(t)$ (diagonal elements of $\tilde{\sigma}(t)$, $p_n(t) = \tilde{\sigma}_{nn}(t)$) decouple from the coherences,\cite{peng2023polaritonic} and thus, for the results shown below, we ignore the coherences and solve the dynamics with the QDs initiated in the second bright excitonic state (located $\approx 70$~meV above the band-edge exciton). $K_{nm,kl}(t)$ is the memory kernel under second-order perturbation~\cite{izmaylov2011nonequilibrium} (See Supporting Information for details). The lattice vibrations were initially in thermal equilibrium (with respect to the ground electronic state) at $T=300$ K. The dynamical pathways described by the model Hamiltonian are illustrated in Figure~\ref{fig:ext_def_spectral}(b).

Fig.~\ref{fig:multi_defect_population} illustrates four distinct dynamical regimes characterized by varying the single-hole depths $\Delta\varepsilon_{\rm ht}$ and diagonal reorganization energies $\lambda_{dd}$ for the lowest defect state. Panels (a)--(c) depict scenarios with a fixed hole depth, $\Delta\varepsilon_{\rm ht}=-0.11\,\mathrm{eV}$, where the equilibrium energies of the defect excitons, $E_{d_n}^0$, remain above the lowest intrinsic exciton levels. When the defect reorganization energy is small (panel (a)), the defect normalized polaron energies, $\mathcal{E}_{d_n} = E_{d_n}^0 - \lambda_{d_nd_n}$, remain above the lowest intrinsic exciton, but shift closer to it due to the relatively small reorganization energy of intrinsic excitons. Specifically, one defect state exhibits a polaron energy around $2.1\,\mathrm{eV}$, while a manifold of additional defect states lie above $2.3\,\mathrm{eV}$, as depicted in the inset of panel (a). In this case, exciton trapping is fast (with a timescale about $1\,\mathrm{ps}^{-1}$), occurring on a time scale comparable to intrinsic hot exciton relaxation.\cite{jasrasaria2023circumventing} However, since the gap of polaron energy between the defect and band-edge excitons is significantly larger than the thermal energy, the equilibrium defect population remains low (as shown by the long-time values of the populations), and the quantum yield remain high with negligible emission from the defect excitons.

When the reorganization energy increases above a certain threshold that depends on the value of the hole trap energy and the spectral density, the polaron energy of the lowest defect state shifts below that of the lowest intrinsic exciton (panel (b)).  In this case, despite having a hole that resides well inside the valance band, trapping becomes slower (with a timescale about $50\,\mathrm{ps}$), resulting from the opening of a gap between the lowest intrinsic and defect polarons.\cite{nitzan1975energy} Despite the slower population of the lowest defect, its equilibrium population is substantially higher than that for small defect reorganization energy, leading to nearly equal steady-state populations.  This suggests that subsequent radiative or nonradiative recombination processes can occur through either the intrinsic exciton states or the defect exciton states, which has been widely reported experimentally.\cite{kambhampati_hot_2011,kambhampati_kinetics_2015,jiang2023effect}

For even larger reorganization energies (panel (c)), additional defect states move into the optical gap, with polaron energies comparable to the lowest defect exciton. As a consequence, the exciton population relaxes rapidly (within approximately $10\,\mathrm{ps}$) through a cascade of relaxation events via multiple defect states. As a result, the intrinsic excitons remain scarcely populated at steady state, and the emission is primarily governed by the lowest defect state.

None of the dynamical regimes shown in panels (a)-(c) of Fig.~\ref{fig:multi_defect_population} exhibit significant nonradiative relaxation to the ground state. In contrast, as illustrated in panel (d), when the hole trap depth increases to $\Delta\varepsilon_{\rm ht}=0.41\,\mathrm{eV}$ (inside the fundamental gap) and for a reorganization energy of $\lambda_{dd}=0.51\,\mathrm{eV}$, a manifold of defect states fills nearly half of the optical gap. Under these conditions, the intrinsic exciton is trapped rapidly, on a timescale of $3\,\mathrm{ps}$, followed by nonradiative relaxation of the defects to the ground state, on a longer timescale of $10\,\mathrm{ns}$. Such rapid trapping, accompanied by relatively efficient nonradiative recombination, is generally undesirable for quantum dot applications.
\begin{figure*}[ht!]
    \centering
    \includegraphics[width=1.0\linewidth]{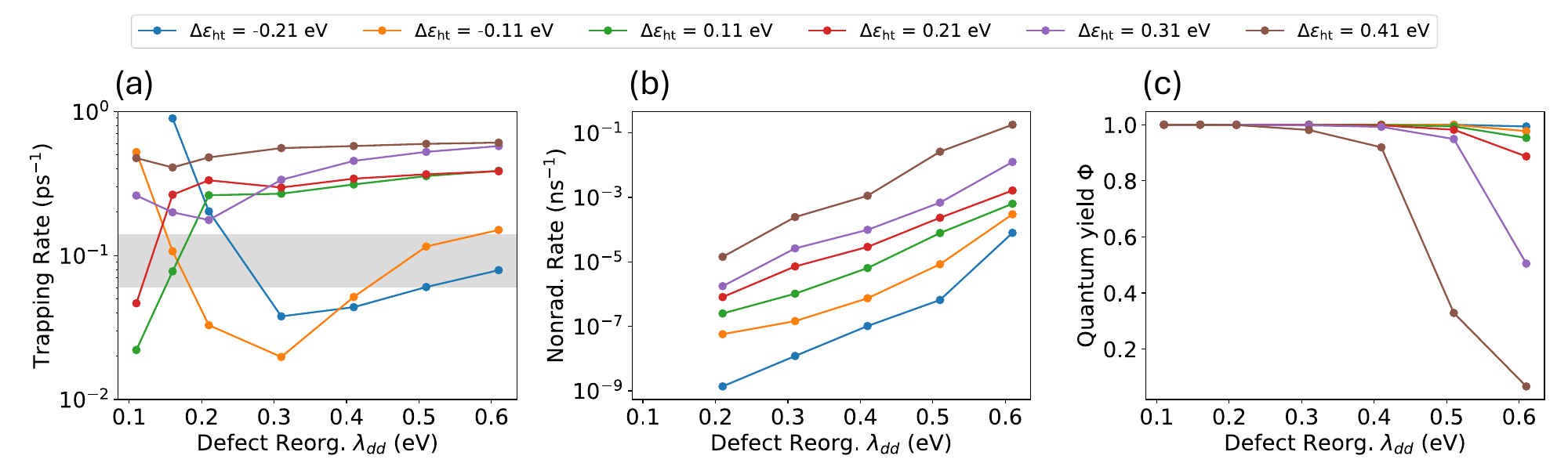}
    \caption{Averaged exciton trapping rate, nonradiative recombination rate, and quantum yield of the QD as functions of defect reorganization energy $\lambda_{dd}$ for systems with varying hole‑trap depths $\Delta\varepsilon_{\rm ht}$.  
    (a) Averaged trapping rate $k_{\rm t}$, extracted by fitting the decay of the excitonic population $1 - \sum_{n\in \rm exct} p_n(t)$ to a single exponential. The shaded region denotes the experimentally measured range of exciton trapping rates for CdSe QDs with diameters of 3.6–6.6 nm.~\cite{sewall_state-resolved_2008}  
    (b) Averaged nonradiative recombination rate $\bar{k}^{(\rm nr)}$, obtained by fitting the ground‑state population decay $1 - p_g(t)$ to a single exponential.  
    (c) Averaged quantum yield $\bar{\Phi}_{\rm qd}$ of the QD, calculated according to Eq.~\eqref{eq:qy}.}
    \label{fig:multi_defect_rates}
\end{figure*}

To systematically analyze exciton trapping and nonradiative recombination over a broad range of defect reorganization energies $\lambda_{dd}$ and hole trap depths $\Delta\varepsilon_{\rm ht}$, we introduce two averaged rates derived from the population dynamics. The average exciton trapping rate, $k_{\rm t}$, is obtained by fitting the decay of excitons population $\sum_{n \in \rm exct} p_n(t)$, to a single exponential function, while the average nonradiative recombination rate, $\bar{k}^{(\rm nr)}$, is determined by fitting the population of the ground state, $1 - p_g(t)$ to a single exponential rise.

The dependence of the trapping rates on the defect reorganization energy for various hole trap depths is shown in Fig.~\ref{fig:multi_defect_rates}(a). For weak to intermediate defect reorganization energies ($\lambda_{dd} < 0.3\,\mathrm{eV}$), the behavior of the rates with respect to the defect reorganization energy depends strongly on the relative positions of the defect polaron energies, $\mathcal{E}_{d_n}$, and the intrinsic polaron energies, $\mathcal{E}_{e_n}$. In systems with hole traps inside the valence band ($\Delta\varepsilon_{\rm ht} = -0.21$ and $-0.11,\mathrm{eV}$), only one defect state lies near and above the lowest intrinsic exciton. In this case, the trapping rate, $k_{\rm t}$, governed by the coupling to a single defect exciton state, decreases with increasing $\lambda_{dd}$. This is primarily because the defect exciton shifts into the optical gap as $\lambda_{dd}$ increases, leading to a larger energy separation from the intrinsic exciton and consequently slower rates.

Conversely, at large reorganization energies ($\lambda_{dd} > 0.3\,\mathrm{eV}$), the trapping rates increase with the defect reorganization energy, regardless of the value of $\Delta\varepsilon_{\rm ht}$. This increase is primarily due to the appearance of additional defect excitonic states near the lowest intrinsic exciton. For small values of $\lambda_{dd}$, these additional states lie above the intrinsic excitonic states but shift down as $\lambda_{dd}$ increases, resulting in faster trapping rates.

Shifting to the averaged nonradiative relaxation rates, plotted in Fig.~\ref{fig:multi_defect_rates} panel (b), we find that they increase monotonically with both $\lambda_{dd}$ and $\Delta\varepsilon_{\rm ht}$. The nonradiative recombination rates are substantially lower than the trapping rates, with the highest rates found in systems containing several in-gap defect excitonic states and strong couplings to the lattice vibrations. Since intrinsic excitons do not relax to the ground state within these timescales, the averaged nonradiative rates primarily reflect decay of the intrinsic exciton through the defect states to the ground state, and thus, the detailed defect spectrum determines the nonradiative rate and the resulting quantum yield, shown in Fig.~\ref{fig:multi_defect_rates}(c). 

The quantum yield is defined as
\begin{equation}\label{eq:qy}
   \bar{\Phi}_{\rm qd} = \frac{k_e^{(\mathrm{r})}}{k_e^{(\mathrm{r})}+k_e^{(\mathrm{nr})}+k_{\rm t}} + \frac{k_{\rm t}}{k_e^{(\mathrm{r})}+k_e^{(\mathrm{nr})}+k_{\rm t}}\cdot \frac{k_d^{(\mathrm{r})}}{k_d^{(\mathrm{r})}+k_d^{(\mathrm{nr})}}
\end{equation}
where $k^{\rm (r)}_e$ ($k^{\rm (nr)}_e$) and $k^{\rm (r)}_d$ ($k^{\rm (nr)}_d$) are the radiative (nonradiative) rates of the intrinsic and defect excitons, respectively (see SI for details). For shallow traps, the quantum yield is nearly unity and decreases slightly with increased reorganization energies of the defects. For deeper traps, the quantum yield is also near unity for small reorganization energies of the defects, but as $\lambda_{dd}$ increases, the quantum yield drops to nearly $10\%$ due to rapid nonradiative recombination.

\begin{figure*}[t]
    \centering
    \includegraphics[width=0.8\linewidth]{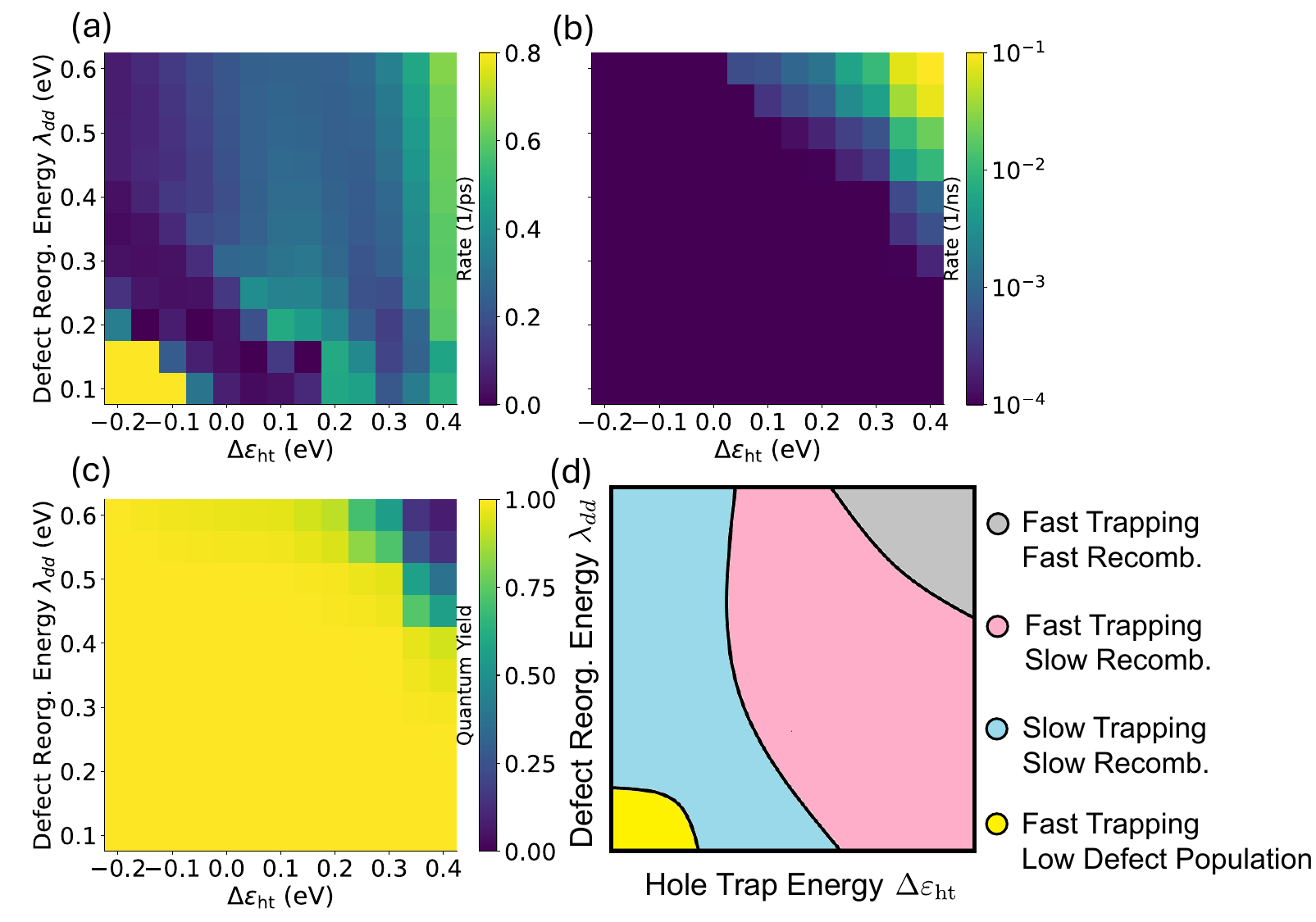}
    \caption{Summary of the multilevel defect–exciton model system. Panels (a)–(c) display (a) the averaged exciton trapping rate $k_{\rm t}$, (b) the averaged nonradiative recombination rate $\bar{k}^{(\rm nr)}$, and (c) the averaged quantum yield $\bar{\Phi}_{\rm qd}$ as functions of defect reorganization energy $\lambda_{dd}$ and hole trap depth $\Delta\varepsilon_{\rm ht}$. Panel (d) presents a schematic diagram of the distinct trapping and recombination regimes.}
    \label{fig:multi_level_model}
\end{figure*}

To summarize the trends observed in the trapping and nonradiative channels, we plot $k_{\rm t}$, $\bar{k}^{(\rm nr)}$, and the averaged quantum yield $\bar{\Phi}_{\rm qd}$ with respect to the two controlled parameters, $\Delta\varepsilon_{\rm ht}$ and $\lambda_{\rm dd}$, in Fig.~\ref{fig:multi_level_model} panels (a)--(c). Panel (d) schematically illustrates distinct regimes of trapping and nonradiative recombination dynamics. As shown in Fig.~\ref{fig:multi_level_model}(a), for hole traps embedded within the valence states ($\Delta\varepsilon_{\rm ht}<-0.1\,\mathrm{eV}$) and weak defect reorganization energies ($\lambda_{dd}<0.2\,\mathrm{eV}$), there is a narrow regime characterized by fast trapping dynamics (colored yellow in panel (d)). As discussed previously in Fig.~\ref{fig:multi_defect_population}(a), this fast trapping rate results from resonances between a single defect state and multiple intrinsic excitonic states. Since the defect polaron energy is above the band-edge exciton, the equilibrium population of the defect state remains small, barely affecting the intrinsic exciton population and quantum yield efficiency. 

The blue region in panel (d) represents a relatively slow trapping regime, defined by the boundary $k_{\rm t}<0.1\,\mathrm{ps}^{-1}$. In this parameter regime, one defect state lies below the intrinsic band-edge exciton, while all other defect states are energetically higher. As a result of an increasing gap between the lowest defect state and the intrinsic excitonic states with increasing trap or reorganization energies, the trapping becomes slower, resulting in a high quantum yield and vanishing nonradiative recombination from the defect state.   

As either $\lambda_{dd}$ or the $\Delta\varepsilon_{\rm ht}$ increase further, the system transitions to a regime of fast trapping, characterized by $k_{\rm t}>0.1\,\mathrm{ps}^{-1}$ (indicated in red in panel (d)). The trapping rate increases monotonically with both $\lambda_{dd}$ and $\Delta\varepsilon_{\rm ht}$ due to the increase in the density of defects above and below the band edge exciton. Although this regime can maintain a high quantum yield, emission spectra may exhibit significant contributions from defects, resulting in a noticeable redshift compared to the intrinsic exciton emission\cite{steenbock2024surface} (the main peak in the photoluminescence spectrum).

For deeper defect levels and larger reorganization energies ($\Delta\varepsilon_{\rm ht}>0.1\,\mathrm{eV}$ and $\lambda_{dd}>0.3\,\mathrm{eV}$), nonradiative transitions between defect excitons and the ground state become significant due to the appearance of mid-gap defect states, stabilized by their reorganization energy. This regime is illustrated in Fig.~\ref{fig:multi_defect_population}(d), gray region. Nonradiative recombination is enhanced due to relaxation through a cascade of defect states. Consequently, the averaged quantum yield $\bar{\Phi}_{\rm qd}$ decreases significantly, approaching zero in the limiting case. This regime, characterized by both rapid trapping and rapid nonradiative recombination, is particularly detrimental to the performance of QD-based applications.

In summary, we presented an atomistic model Hamiltonian, parametrized by a semiempirical pseudopotential approach, to investigate the mechanisms and timescales of exciton trapping and nonradiative recombination in CdSe/CdS core/shell quantum dots (QDs) containing a single surface hole trap. By systematically varying the hole trap depth and defect reorganization energy (polaron shift), we have mapped out the interplay between intrinsic and defect exciton dynamics across a broad parameter space. Our study reveals that a single unpassivated surface sulfur atom can generate multiple defect excitonic states, some of which lie within or below the optical gap. The relative polaron energies between intrinsic and defect excitons govern the efficiency of exciton trapping and nonradiative decay. We identify distinct dynamical regimes: slow trapping when isolated defect states are energetically detuned from intrinsic excitons, and fast trapping when resonance conditions with a dense manifold of defect states are met. Furthermore, we uncover a mechanism for enhanced nonradiative recombination via cascade relaxation through in-gap defect states. These insights highlight the critical role of surface defect energetics and vibronic couplings in determining QD photophysics, offering design principles for tailoring quantum yield through improved surface passivation and defect engineering strategies.

\section*{Acknowledgments}
We thank Matthew Coley‑O’Rourke, Eric R. Heller, and Kritanjan Polley for valuable discussions. This work was supported by the U.S. Department of Energy, Office of Science, Office of Basic Energy Sciences, Materials Sciences and Engineering Division, under Contract No. DE-AC02-05CH11231, as part of the Fundamentals of Semiconductor Nanowire Program (KCPY23). This research used resources of the National Energy Research Scientific Computing Center, a DOE Office of Science User Facility supported by the Office of Science of the U.S. Department of Energy under Contract No. DE-AC02-05CH11231 using NERSC award BES-ERCAP0032503. S.G acknowledges the support from the European Union’s Framework Programme for Research and Innovation Horizon 2020 (2014-2020) under the Marie Skłodowska-Curie Grant Agreement No. 847471. S.L.R. acknowledges support from the Alexander von Humboldt Foundation via the award of a research fellowship.

\section*{Reference}

\bibliographystyle{apsrev}
\bibliography{main}

\end{document}


\oldmaketitle

\section{QD Configuration}
The CdSe-CdS core-shell colloidal quantum dot (QD) monomer was constructed by adding CdS shells to a faceted CdSe core seed with a lattice constant of bulk wurtzite CdSe ($a=4.3$ \AA, $c=\sqrt{\frac{8}{3}}a$). Each QD has a diameter of 3.5 nm with a CdSe core of 2.0 nm (Cd$_{798}$Se$_{108}$S$_{690}$).The structures were minimized with Stillinger-Weber force field parameterized for II-VI nanostructures implemented in LAMMPS~\cite{plimpton_fast_1995,zhou_stillinger-weber_2013}. The defect state is introduced by removing one of the passivation ligands of sulfur atom on [001] plane.

\section{Pseudopotential Calculation}
The semi-empirical pseudopotential model was employed to calculate the quasi-electron and hole states of the QD NC. The local screened pseudopotentials were fitted to a functional form in the reciprocal space as~\cite{jasrasaria2021interplay}
\begin{equation}\label{eq:pseudo}
\Tilde{v}(q)=a_0\left[1+a_4 \operatorname{Tr} \bm{\epsilon}\right] \frac{q^2-a_1}{a_2 \exp \left(a_3 q^2\right)-1}
\end{equation}
where $q$ is the magnitude of momentum vector, $\bm{\epsilon}$ is the strain tensor, and $a_0$ to $a_4$ are fitting parameters for Cd, Se, and S, which are tabulated in the Supplementary Table~1. The parameters were fitted simultaneously to the bulk band structures and deformation potentials of CdSe and CdS to capture the electronic and vibronic properties of the nanocrystals. The real-space quasi-electron Hamiltonian ${h}_{\rm e}(\mathbf{r})$ was given by~\cite{cohen1967,wang1995local,fu_inp_1997}
\begin{equation}
    {h}_{\rm e}(\mathbf{r}) = -\frac{1}{2}\nabla^2_{\mathbf{r}} + \sum_\mu v_\mu\left(|\mathbf{r}-\mathbf{R}_{0,\mu}|\right),
\end{equation}
where $v_\mu(\mathbf{r})$ is the real-space pseudopotential for atom $\mu$, obtained from the Fourier transform of $\Tilde{v}(q)$. Filter diagonalization technique was then applied to obtain quasi-electron states $\psi_i(\mathbf{r})$ above the conduction band edge. The calculations were performed on real-space grids smaller than 0.7 a.u., ensuring eigenenergies converged to less than $10^{-3}$ meV, and sequential calculations of electron-phonon couplings were converged.
\begin{table}[ht!]
\begin{tabular}{|c|c|c|c|c|c|}
\hline
   & $a_0$    & $a_1$       & $a_2$     & $a_3$      & $a_4$     \\ \hline
Cd & -31.4518 & 1.3890 & -0.0502 & 1.6603 & 0.0586 \\ \hline
Se & 8.4921 & 4.3513   & 1.3600 & 0.3227  & 0.1746 \\ \hline
S  & 7.6697 & 4.5192   & 1.3456 & 0.3035  & 0.2087 \\ \hline
\end{tabular}
\caption{Pseudopotential parameters in Eq.~\eqref{eq:pseudo} for Cd, Se, and S atoms. All parameters are given in atomic units.}
\label{tab:S1}
\end{table}

\section{Bethe–Salpeter Equation}
The exciton states are represented as linear combinations of non-interacting electron and hole states~\cite{rohlfing2000electron,eshet2013electronic}
\begin{equation}\label{eq:psi_n}
    \left|\psi_n\right\rangle=\sum_{\substack{{a\in\rm{elec}} \\ {i\in\rm{hole}}}} c_{ai}^n {d_a^{\dagger}} {d_i}|\psi_g\rangle
\end{equation}
where $d^{\dagger}$ ($d$) are fermionic creation (annihilation) operators, $a$, $i$ label the electron and hole state respectively, and $n$ labels (defect or intrinsic) excitonic states. The BSE coefficient $c_{ai}^n$ is determined from  Bethe-Salpeter equation (BSE)
\begin{equation}
    \left(E_n^0-\Delta \varepsilon_{a i}\right) c_{a, i}^n=\sum_{\substack{b\in \rm{elec}\\j \in \rm{hole}}}\left(K_{a i ; b j}^d+K_{a i ; b j}^x\right) c_{b, j}^n
\end{equation}
where $\Delta\varepsilon_{ai}=\varepsilon_a-\varepsilon_i$. $K_{a i ; b j}^d$ and $K_{a i ; b j}^x$ are the electron-hole kernels for the screened direct Coulomb attraction and exchange interaction. In the calculation, 64 electron states and 100 hole states are used to converge the exciton energies.

\section{Phonon Couplings of Intrinsic and Defect Excitons}
In the crude adiabatic approximation, the exciton couplings in the atomic coordinates $\mathbf{R}$ relative to the ground state are denoted as
\begin{equation}\label{eq:Vnm}
    V_{nm}^{\mu k} \equiv \left\langle\psi_{n}\left|\left(\frac{\partial v_\mu}{\partial {R}_{\mu k}}\right)_{\mathbf{R}_{0}}\right| \psi_{m}\right\rangle - \delta_{nm}\left\langle\psi^g\left|\left(\frac{\partial v_\mu}{\partial {R}_{\mu k}}\right)_{\mathbf{R}_{0}}\right| \psi^g\right\rangle,
\end{equation}
where the derivatives of the pseudopotential with respect to the nuclear coordinate $\left(\frac{\partial v_\mu}{\partial {R}_{\mu k}}\right)_{\mathbf{R}_{0}}$ is taken at the equilibrium configuration $\mathbf{R}_0$. The pseudopotential derivative operator can be written in the 2nd quantization as
\begin{equation}\label{eq:dvdR_sqz}
    \frac{\partial v_\mu\left(\left|\mathbf{r}-\mathbf{R}_\mu\right|\right)}{\partial R_{\mu k}}\equiv\sum_{r s} v_{r s, \mu k}^{\prime}\left(R_{\mu k}\right) d_r^{\dagger} d_s,
\end{equation}
where indices $r$ and $s$ run over all the electron and hole states. Using Eqs.~\eqref{eq:psi_n}, \eqref{eq:Vnm} and \eqref{eq:dvdR_sqz}, the couplings between two excitonic states $V_{nm}^{\mu k}$ in the atomic corrdinates can then be written as~\cite{jasrasaria2021interplay}
\begin{equation}
\begin{aligned}
    V_{nm}^{\mu k} &= \left\langle\psi_{n}\left|\left(\frac{\partial v_\mu}{\partial {R}_{\mu k}}\right)_{\mathbf{R}_{0}}\right| \psi_{m}\right\rangle - \delta_{nm}\left\langle\psi^g\left|\left(\frac{\partial v_\mu}{\partial {R}_{\mu k}}\right)_{\mathbf{R}_{0}}\right| \psi^g\right\rangle\\
    &=\sum_{\substack{ab\in \rm{elec}\\ij\in \rm{hole}}} c_{a, i}^n c_{b, j}^m \sum_{r s} v_{r s, \mu k}^{\prime}\left(R_{\mu k}\right)\left\langle \psi^g\left|d_i^{\dagger} d_a d_r^{\dagger} d_s d_b^{\dagger} d_j\right| \psi^g\right\rangle-\delta_{nm}\sum_{i\in {\rm \rm{hole}}}v_{ii, \mu k}^{\prime}\left(R_{\mu k}\right)\\
     &=\sum_{\substack{ab\in \rm{elec}\\i\in \rm{hole}}} c_{a, i}^n c_{b, i}^m v_{a b, \mu k}^{\prime}\left(R_{\mu k}\right)-\sum_{\substack{a\in \rm{elec}\\ij\in \rm{hole}}} c_{a, i}^n c_{a, j}^m v_{i j, \mu k}^{\prime}\left(R_{\mu k}\right),
\end{aligned}
\end{equation}
where $a, b$ indices are summed over electron states, $i, j$ indices are summed over hole states and atomic coordinates label $\mu\in1\dots N_{\rm atoms}$, $k\in x,y,z$. Similarly, the couplings between the excitonic states to the ground state are
\begin{align}
    V_{ng}^{\mu k}&=\bra{\psi_n}\sum_{r s} v_{r s, \mu k}^{\prime}\left(R_{\mu k}\right) d_r^{\dagger} d_s\ket{\psi^g} = \sum_{\substack{a\in \rm{elec}\\i\in \rm{hole}}}\sum_{r s} c_{a,i}^{n*} v_{r s, \mu k}^{\prime}\left(R_{\mu k}\right) \bra{\psi_g}d_i^\dagger d_a d_r^{\dagger} d_s\ket{\psi_g}\\
    &=\sum_{\substack{a\in \rm{elec}\\i\in \rm{hole}}}\sum_{r s} c_{a,i}^{n*} v_{r s, \mu k}^{\prime}\left(R_{\mu k}\right)\delta_{ar}\delta_{is} = \sum_{\substack{a\in \rm{elec}\\i\in \rm{hole}}}c_{a,i}^{n*}v_{a i, \mu k}^{\prime}\left(R_{\mu k}\right).
\end{align}
The transformation between normal mode and atomic coordinate is $Q_\alpha  = \sum_{\mu k}\sqrt{m_\mu} E_{\mu k}^\alpha(R_{\mu k}-R_{\mu k,0})$, which leads to the definition of electron-phonon couplings
\begin{equation}
    V_{nm}^\alpha = \sum_{\mu,k}\frac{1}{\sqrt{m_\mu}} E_{\mu k,\alpha}V_{nm}^{\mu k}, \hspace{0.5cm} \alpha\in1\dots 3N_{\rm atoms}-6.
\end{equation}
The spectral density characterizing the structure of exciton-phonon couplings is defined as 
\begin{align}
\begin{aligned}
    J_{nm}(\omega) &= \pi\sum_\alpha\frac{\left(V_{nm}^\alpha\right)^2}{2\omega_\alpha}\delta(\omega-\omega_\alpha) = \pi\sum_\alpha\omega_\alpha\lambda^\alpha_{nm}\delta(\omega-\omega_\alpha)
\end{aligned}
\end{align}
where the mode-wise reorganization energy $\lambda^\alpha_{nm}=\frac{\left(V_{nm}^\alpha\right)^2}{2\omega_\alpha^2}$ and total reorganization energy is defined as $\lambda_{nm}=\sum_\alpha\lambda^\alpha_{nm}$. The spectral density is normalized as $\lambda_{nm}=\frac{1}{\pi} \int d\omega \frac{J_{nm}(\omega)}{\omega}$. Figure~2 in the main text shows the spectral density with broadening of the Delta function as $\delta(\omega-\omega_\alpha)\approx\frac{1}{\sigma \sqrt{2\pi}}e^{-\frac{(\omega-\omega_\alpha)^2}{2\sigma^2}}$ and the broadening factor $\sigma=0.1$ meV. 

\section{Model Hamiltonian}
Based on the BSE and calculation of exciton-phonon couplings, we can write out the model Hamiltonian for QDs as
 \begin{align}
 \begin{aligned}
    & H_{\rm QD} = E_g\ket{\psi_g}\bra{\psi_g} +\sum_{\alpha}\left(\frac{1}{2}{P}_{\alpha}^2 + \frac{1}{2}\omega_\alpha^2Q_\alpha^2\right) \\
    & + \sum_{n, m\in \rm{exct}} \left(E_{e_n}^0\delta_{nm} + \sum_\alpha V_{e_ne_m}^\alpha Q_\alpha\right)\ket{\psi_{e_n}}\bra{\psi_{e_m}}+ \sum_{\substack{n\in\text{exct}\\\alpha}}\ket{\psi_{e_n}}\bra{\psi_{g}} V_{e_ng}^\alpha Q_\alpha+h.c.\\
    & + \sum_{n, m\in \text{def}}\left(E_{d_n}^{0}\delta_{nm}+\sum_{\alpha}\mathcal{V}_{d_nd_m}^\alpha{Q}_{\alpha}\right)\ket{\psi_{d_n}}\bra{\psi_{d_m}}\\
    & +  \sum_{\substack{n\in\text{def}, \\m\in\text{exct}\\\alpha}} \ket{\psi_{d_n}}\bra{\psi_{e_m}}\mathcal{V}_{d_ne_m}^\alpha{Q}_{\alpha} + \sum_{\substack{n\in\text{def}\\\alpha}} \ket{\psi_{d_n}}\bra{\psi_g}\mathcal{V}_{{d_n}g}^\alpha{Q}_{\alpha} + h.c.
 \end{aligned}
 \label{eq:hamiltonian}
\end{align}
where $E_g$ ($E_g=0$ as a reference), $E_{e_n}^0$ and $E_{d_n}^0$ are energies of ground state, excitonic states and defect state at \textit{equilibrium} configuration ($\mathbf{Q}=0$), calculated from BSE. $V_{d_nd_m}^\alpha$ are intrinsic exciton-phonon couplings among excitonic states, and $V_{d_ng}^\alpha$ are intrinsic exciton-phonon couplings between one intrinsic exciton state and the ground state. Those couplings are not scales since they have been carefully parameterized and tested. $\mathcal{V}_{d_nd_m}^\alpha$ are vibronic couplings that involve at least one defect state and are scaled according to the following scheme. We scale the diagonal phonon coupling of the defect by $a\mathcal{V}_{d_nd_n}^\alpha$ and off-diagonal coupling by $b\mathcal{V}_{d_n\neq d_m}^\alpha$, where ($a,b<$1, $n\in \rm{defect}$, $m\in \rm{exciton}$ or ground state). Note that for the reorganization energy $\lambda_{d_nd_n}^{\rm scale} = a^2\lambda_{d_nd_n}$ and $\lambda_{d_n\neq d_m}^{\rm scale} = b^2\lambda_{d_nd_m}$. We want to control the scaled reorganization energy for the \textbf{lowest defect state} to be $\lambda_{dd}\equiv\lambda_{d_1d_1}^{\rm scale} = 0.1-0.6$ eV and off-diagonal with \textit{the lowest exciton state} to be $\lambda_{ed}\equiv\lambda_{e_1d_1}^{\rm scale} = 0.1$ meV.

\section{Polaron-Transformed Quantum Master Equation}
Before deriving the quantum master equation, we first simplify the notation of the model Hamiltonian
\begin{align}
    H_{\rm QD} &=  H_{\rm ex} + H_{\rm nu} + H_{\rm ex-nu}\\
    H_{\rm ex} &= \sum_n E_n\ket{\psi_n}\bra{\psi_n}\\
    H_{\rm nu} &= \sum_\alpha\frac{P_\alpha^2}{2} + \frac{1}{2}\omega_\alpha^2Q_\alpha^2\\
    H_{\rm ex-nu} &= \sum_{nm}\ket{\psi_n}\bra{\psi_m}\sum_\alpha V_{nm}^\alpha Q_\alpha
\end{align}
where we suppress the labels for intrinsic and defect excitons. To treat the strong diagonal exciton-phonon couplings, we transform the original Hamiltonian into the polaron-transformed framework
\begin{equation*}
\begin{split}
    \Tilde{H}_{\rm QD} &=  \sum_{n}\mathcal{E}_n\ket{n}\bra{n} + \sum_\alpha\frac{P_\alpha^2}{2} + \frac{1}{2}\omega_\alpha^2Q_\alpha^2 +  \sum_{n\neq m} \ket{n}\bra{m} e^{S_n} \sum_\alpha V_{nm}^\alpha Q_\alpha e^{-S_m}\\
    &= \Tilde{H}_{\rm ex} + {H}_{\rm nu} + \Tilde{H}_{\rm ex-nu}
\end{split}
\end{equation*}
where the polaron energy $\mathcal{E}_n = E_n-\lambda_n$, and the displaced operator $e^{S_n} = \exp{-\frac{i}{\hbar}\sum_{\alpha}\frac{V_{nn}^\alpha}{\omega_\alpha^2} P_\alpha}$. We define the zero-th order Hamiltonian $H_0 = \Tilde{H}_{\rm el} + {H}_{\rm nu} + \left<\Tilde{H}_{\rm el-nu}\right>_{\rm nu}$, and the perturbation $H' = \Tilde{H}_{\rm el-nu} -  \left<\Tilde{H}_{\rm el-nu}\right>_{\rm nu}$. Note that $\left<\Tilde{H}_{\rm el-nu}\right>_{\rm nu}\approx0$ in the full polaron transformation framework.\cite{jang2022partially} The dynamics of the reduced density matrix $\Tilde{\sigma}(t)$ under the second-order approximation to the memory kernel is given by
\begin{align}\label{eq:pt_eom}
\frac{\partial \tilde{\sigma}_{nm}(t)}{\partial t} 
&= -i \tilde{\omega}_{nm}\,\tilde{\sigma}_{nm}(t) + \sum_{kl} K_{nm,kl}(t)\,\tilde{\sigma}_{kl}(t).
\end{align}
where the approximated memory kernel can be written as
\begin{equation}
    K_{nm,kl}(t) = R_{lm,nk}(t, \omega_{kn}) + R^*_{kn,ml}(t, \omega_{lm}) -\delta_{lm}\sum_rR_{nr,rk}(t, \omega_{kr}) - \delta_{nk}\sum_r R^*_{mr,rl}(t,\omega_{lr}).
\end{equation}
The Redfield tensor 
\begin{equation}
    R_{nm, kl}(t, \omega) = \frac{1}{\hbar^2} \int_0^t d\tau C_{nm,kl}(\tau)e^{i\omega \tau},
\end{equation}
and the correlation function can be written analytically as\cite{izmaylov2011nonequilibrium}
\begin{align}
\begin{aligned}
        C_{nm,kl}(t) &= \left[h_{nm,kl}(t)h_{lk,nm}(t) + l_{nm,kl}(t) \right]f_{nm,kl}^{\rm FC}(t)\\
        h_{nm,kl}(t) &= \sum_\alpha-\frac{{V}_{nm}^\alpha}{2\omega^2_\alpha}\left({V}_{kk}^\alpha-{V}_{ll}^\alpha\right)\left[n_\alpha e^{i\omega_\alpha t}-(n_\alpha+1) e^{-i\omega_\alpha t}\right]-\frac{{V}_{nm}^\alpha}{2\omega^2_\alpha}\left({V}_{nn}^\alpha+{V}_{mm}^\alpha\right)\\
    l_{nm,kl}(t) &= \frac{\hbar{V}_{nm}^\alpha{V}_{kl}^\alpha }{2\omega_\alpha}\left[n_\alpha e^{i\omega_\alpha t} + (n_\alpha+1) e^{-i\omega_\alpha t}\right] \\
    f_{nm,kl}^{\rm FC}(t) &= \exp \Bigg\{ \sum_\alpha\frac{1}{2\hbar\omega_\alpha^3}\left({V}_{mm}^\alpha - {V}_{nn}^\alpha\right)\left({V}_{kk}^\alpha - {V}_{ll}^\alpha\right) \left[\left({n}_\alpha + 1\right) e^{-i \omega_\alpha t} + n_\alpha e^{i \omega_\alpha t}\right] \\
& \hspace{1cm} - \frac{1}{2\hbar\omega_\alpha^3}\left[\left({V}_{mm}^\alpha - {V}_{nn}^\alpha\right)^2 + \left({V}_{kk}^\alpha - {V}_{ll}^\alpha\right)^2\right]\left(n_\alpha+\frac{1}{2}\right)\Bigg\}.
\end{aligned}
\end{align}
where the average thermal occupation $n_\alpha=1/(e^{\hbar\omega_\alpha/(kT)}-1)$.

\section{Multi-Defect Population}
\begin{figure}[ht!]
    \centering
    \includegraphics[width=1.0\linewidth]{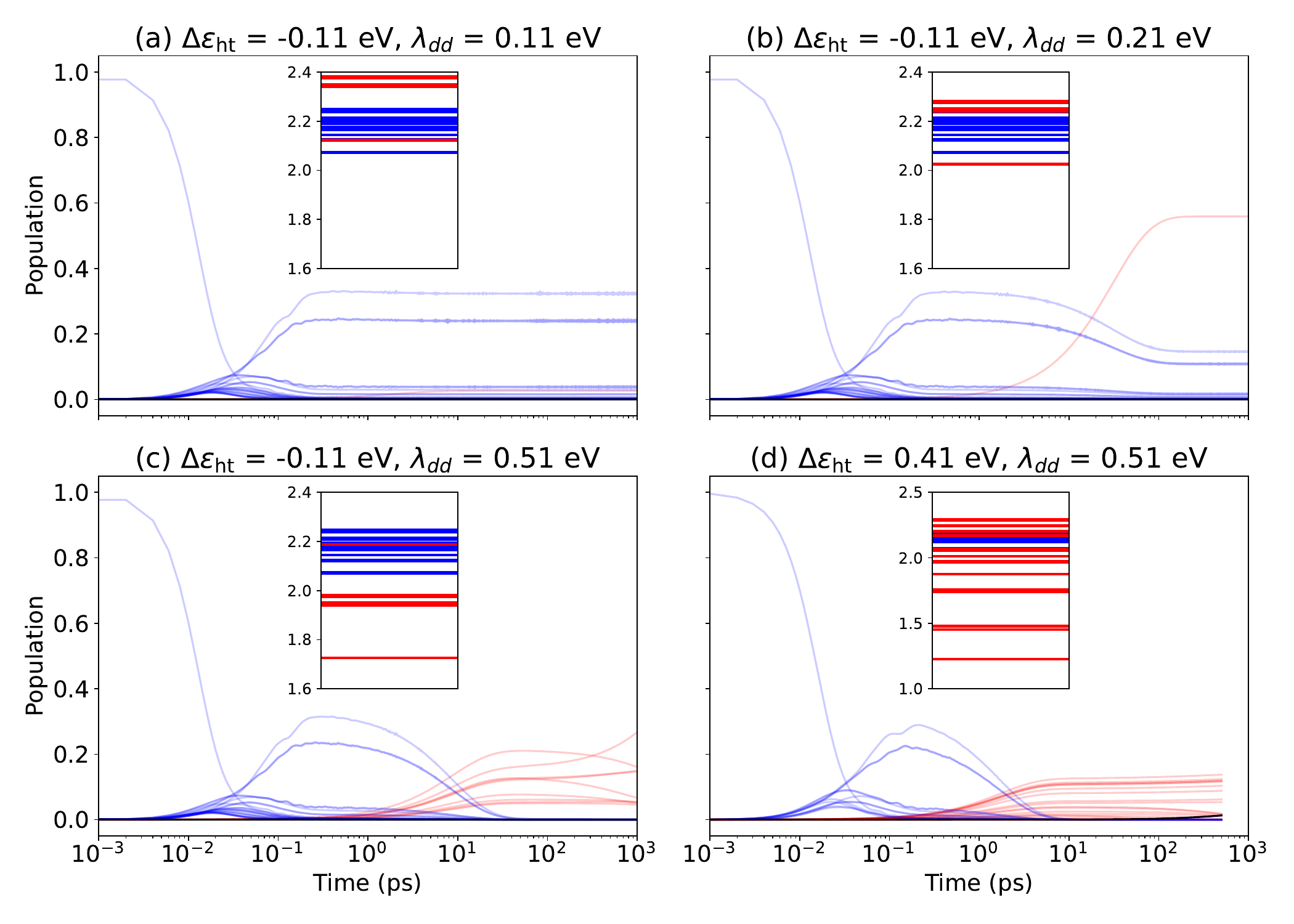}
    \caption{Exciton dynamics in the presence of multiple defect states. Each intrinsic exciton (blue) and defect (red), and ground states (black) are plotted. }
    \label{fig:multi_defect_population}
\end{figure}
Figure~\ref{fig:multi_defect_population} shows the individual multi-defect population dynamics corresponding to Figure~3 in the main text. 

\newpage
\section{Averaged quantum yield}

For the intrinsic exciton states, define the averaged radiative recombination rate $k_e^{(\mathrm{r})}$, averaged nonradiative recombination rate $k_e^{(\mathrm{nr})}$ and the rate of trapping into the defect state $k_{\rm t}$. The probability for direct radiative decay from exciton states is
\[
P_{\mathrm{rad}}^e = \frac{k_e^{(\mathrm{r})}}{k_e^{(\mathrm{r})} + k_e^{(\mathrm{nr})} + k_{\rm t}},
\]
and the probability for trapping is
\[
P_{\rm t} = \frac{k_{\rm t}}{k_e^{(\mathrm{r})} + k_e^{(\mathrm{nr})} + k_{\rm t}}.
\]
The defect states decay radiatively with rate \(k_d^{(\mathrm{r})}\) and nonradiatively with rate \(k_d^{(\mathrm{nr})}\). Hence, the radiative quantum yield in for defects is 
\[
P_{\mathrm{rad}}^d = \frac{k_d^{(\mathrm{r})}}{k_d^{(\mathrm{r})}+k_d^{(\mathrm{nr})}}.
\]
The averaged quantum yield can be calculated by the sum of the two pathways
\begin{align*}
    \bar{\Phi}_{\rm qd} &= P_{\mathrm{rad}}^e + P_{\rm t}\cdot P_{\mathrm{rad}}^d\\
    &= \frac{k_e^{(\mathrm{r})}}{k_e^{(\mathrm{r})}+k_e^{(\mathrm{nr})}+k_{\rm t}} + \frac{k_{\rm t}}{k_e^{(\mathrm{r})}+k_e^{(\mathrm{nr})}+k_{\rm t}}\cdot \frac{k_d^{(\mathrm{r})}}{k_d^{(\mathrm{r})}+k_d^{(\mathrm{nr})}}
\end{align*}
This derivation assumes first-order kinetics and ignores detrapping. Note that $k_d^{(\mathrm{r})}\ll k_d^{(\mathrm{nr})}$ and $k_e^{(\mathrm{r})}\gg k_e^{(\mathrm{nr})}$.

\bibliography{supplement}